# An Extended Result on the Optimal Estimation under Minimum Error Entropy Criterion

*Badong Chen[1], Guangmin Wang[1], Nanning Zheng[1], José C. Príncipe[2]*

1. Institute of Artificial Intelligence and Robotics, Xi'an Jiaotong University, Xi'an 710049, China (chenbd@mail.xjtu.edu.cn)
2. Department of ECE, University of Florida, Gainesville, FL 32611 USA

*Abstract*— **The minimum error entropy (MEE) criterion has been successfully used in fields such as parameter estimation, system identification and the supervised machine learning. There is in general no explicit expression for the optimal MEE estimate unless some constraints on the conditional distribution are imposed. A recent paper has proved that if the conditional density is conditionally symmetric and unimodal (CSUM), then the optimal MEE estimate (with Shannon entropy) equals the conditional median. In this study, we extend this result to the generalized MEE estimation where the optimality criterion is the Renyi entropy or equivalently, the $\alpha$ -order information potential (IP).**

I. INTRODUCTION

Given two random variables: $X \in \mathbb{R}^n$ which is an unknown parameter to be estimated, and $Y \in \mathbb{R}^m$ which is the observation or measurement. The estimation of $X$ based on $Y$, is in general a measurable function of $Y$, denoted by $\hat{X} = g(Y) \in G$, where $G$ stands for the collection of all Borel measurable functions with respect to the $\sigma$-field generated by $Y$. The optimal estimate $g^*(Y)$ can be determined by minimizing a certain risk, which is usually a function of the error distribution. If $X$ has conditional probability density function (PDF) $p(x|y)$, then

$$g^* = \arg\min_{g \in G} \mathcal{R}\left(p^g(x)\right) \tag{1}$$

where $p^g(x)$ is the PDF of the estimate error $E = X - g(Y)$, $\mathcal{R}(.)$ is the risk function: $\mathbb{P}_E \to \mathbb{R}$, $\mathbb{P}_E$ denotes the collection of all possible PDFs of the error. Let $F(y)$ be the distribution function of $Y$, the PDF $p^g(x)$ will be

$$p^g(x) = \int_{\mathbb{R}^m} p(x + g(y) \mid y) dF(y) \tag{2}$$

As one can see from (2), the problem of choosing an optimal $g$ is actually the problem of shifting the components of a mixture of the conditional PDF so as to minimize the risk $\mathcal{R}$.

The risk function $\mathcal{R}$ plays a central role in estimation related problems since it determines the performance surface and hence governs the optimal solution and the performance of the search algorithms. Traditional Bayes risk functions are, in general, defined as the expected value of a certain loss function (usually a nonlinear mapping) of the error:

$$\mathcal{R}_{Bayes}\left(p^g(x)\right) = \int_{\mathbb{R}^n} l(x) p^g(x) dx \tag{3}$$

where $l(.)$ is the loss function. The most common Bayes risk function used for estimation is the mean square error (MSE), also called the squared error or quadratic error risk, which is defined by $\mathcal{R}_{MSE}\left(p^g(x)\right) = \int_{\mathbb{R}^n} \|x\|_2^2 p^g(x) dx$ *. Using the MSE as risk, the optimal estimate of $X$ is simply the conditional mean $\pi(y) \triangleq \text{mean}\left[p(.|y)\right]$. The popularity of the MSE is due to its simplicity and optimality for linear Gaussian case [8, 9, 11]. However, MSE is not always a superior risk function especially for non-linear and non-Gaussian situations, since it only takes into account the second order statistics. Therefore, many alternative Bayes risk functions have been used in practical applications. The mean absolute deviation (MAD) $\mathcal{R}_{MAD}\left(p^g(x)\right) = \int_{\mathbb{R}^n} \|x\|_1 p^g(x) dx$, with which the optimal estimate is the conditional median $\mu(y) \triangleq \text{median}\left[p(.|y)\right]^{\dagger}$, is a robust risk function and has been successfully used in adaptive filtering in impulsive noise environments [16]. The mean 0-1 loss $\mathscr{R}_{0-1}\left(p^g(x)\right) = \int_{\mathbb{R}^n} l_{0-1}(x) p^g(x) dx$, where $l_{0-1}(.)$ denotes the 0-1 loss function$^{\ddagger}$, yields the optimal estimate as $\zeta(y) \triangleq \text{mode}\left[p(.|y)\right]$, i.e. the conditional mode$^{\S}$, which is also the maximum *a posteriori* (MAP)$^{**}$ estimate if regarding $p(.|y)$ as the posterior density. Other important Bayes risk functions include the mean $p$-power error [12], Huber's M-estimation cost [15], and the risk-sensitive cost [1], etc. For general Bayes risk (3), there is no explicit expression for the optimal estimate unless some conditions on $l(x)$ or/and conditional density $p(x|y)$ are imposed. As shown in [6], if $l(x)$ is even and convex, and the conditional density $p(x|y)$ is symmetric in $x$, the optimal estimate will be the conditional mean (or equivalently, the conditional median).

Besides the traditional Bayes risk functions, the error entropy (EE) can also be used as a risk function in estimation problems. Using Shannon's definition of entropy [3], the EE risk function is

$$\mathscr{R}_S\left(p^g(x)\right) = -\int_{\mathbb{R}^n} p^g(x) \log p^g(x) dx \qquad (4)$$

As the entropy measures the average dispersion or uncertainty of a random variable, its minimization makes the error concentrated. Different from conventional Bayes risks, the "loss function" of the EE risk (4) is $-\log p^g(x)$, which is directly related to the error's PDF. Therefore, when using the EE risk, we are nonlinearly transforming the error by its own PDF. In 1970, Weidemann and Stear published a paper entitled "Entropy Analysis of Estimating Systems" [18] in which they studied the parameter estimation problem using the error entropy as a criterion functional. They proved minimizing the error entropy is equivalent to minimizing the mutual information between error and observation, and also proved the reduced error entropy is upper-bounded by the amount of information obtained by observation. Later, Tomita et al [17] and Kalata and Priemer [10] studied the estimation and filtering problems from the viewpoint of the information theory and derived the Kalman filter as a minimum-error-entropy (MEE) linear estimator. Like most Bayes risks, the EE risk (4) has no explicit expression for the optimal estimate unless some constraints on the conditional density $p(x|y)$ are imposed. In a recent paper [2], Chen and Geman proved that, if $p(x|y)$ is conditionally symmetric and unimodal (CSUM), the MEE estimate (the optimal estimate under EE risk) will be the conditional median (or equivalently, the conditional mean or mode). Table 1 gives a summary of the optimal estimates for several risk functions.

---

\* In this paper, $\|.\|_p$ denotes the $p$-norm.

† Here the median of a random vector is defined as the element-wise median vector.

‡ The 0-1 loss function $l_{0-1}(x)$ has been frequently used in statistics and decision theory. If the error is a discrete variable, $l_{0-1}(x) = \mathbb{I}\left(x \neq \mathbf{0}\right)$, where $\mathbb{I}(.)$ is the indicator function, whereas if the error is a continuous variable, $l_{0-1}(x)$ is defined as $l_{0-1}(x) = 1 - \delta(x)$, where $\delta(.)$ is the Dirac delta function.

§ The mode of a continuous probability distribution is the value at which its PDF attains its maximum value.

\*\* The MAP estimate is a limit of Bayes estimator (under the 0-1 loss function), but generally not a Bayes estimator.

| Risk function | $\mathcal{R}(p^g(x))$ | Optimal solution |
|---|---|---|
| Mean square error (MSE) | $\int_{\mathbb{R}^n} \|x\|_2^2 \, p^g(x) dx$ | $g^*(y) = \pi(y) \triangleq \text{mean}\left[p(.|y)\right]$ |
| Mean absolute deviation (MAD) | $\int_{\mathbb{R}^n} \|x\|_1 \, p^g(x) dx$ | $g^*(y) = \mu(y) \triangleq \text{median}\left[p(.|y)\right]$ |
| Mean 0-1 loss | $\int_{\mathbb{R}^n} l_{0-1}(x) p^g(x) dx$ | $g^*(y) = \zeta(y) \triangleq \text{mode}\left[p(.|y)\right]$ |
| General Bayes risk | $\int_{\mathbb{R}^n} l(x) p^g(x) dx$ | If $l(x)$ is even and convex, and $p(x|y)$ is symmetric in $x$, then $g^*(y) = \pi(y) = \mu(y)$ |
| Error entropy (EE) | $-\int_{\mathbb{R}^n} p^g(x) \log p^g(x) dx$ | If $p(x|y)$ is CSUM, then $g^*(y) = \pi(y) = \mu(y) = \zeta(y)$ |

Table 1. Optimal estimates for several risk functions.

4In statistical information theory, there are many extensions to Shannon's original definition of entropy. Renyi's entropy is one of the parametrically extended entropies. Given a random variable $X$ with PDF $p(x)$, $\alpha$-order Renyi entropy is defined by [14]

$$H_\alpha(X) = \frac{1}{1-\alpha} \log\left(\int_{\mathbb{R}^n} (p(x))^\alpha \, dx\right) \tag{5}$$

where $\alpha > 0$, and $\alpha \neq 1$. The entropy definition (5) becomes the usual Shannon entropy as $\alpha \to 1$. Renyi entropy can be used to define a generalized EE risk:

$$\mathcal{R}_\alpha\left(p^g(x)\right) = \frac{1}{1-\alpha} \log\left(\int_{\mathbb{R}^n} \left(p^g(x)\right)^\alpha dx\right) \tag{6}$$

In recent years, the EE risk (6) has been successfully used as an adaptation cost in information theoretic learning (ITL) [4, 5, 13]. It has been shown that the nonparametric kernel (Parzen window) estimator of Renyi entropy (especially when $\alpha = 2$) is more computationally efficient than that of Shannon entropy [13]. The argument of the logarithm in Renyi entropy, denoted by $V_\alpha$ ($V_\alpha = \int_{\mathbb{R}^n} (p(x))^\alpha \, dx$), is called the $\alpha$-*order information potential* (IP)[††] [13]. As the logarithm is a monotonic function, the minimization of Renyi entropy is equivalent to the minimization (when $\alpha < 1$) or maximization (when $\alpha > 1$) of information potential. In practical applications, information potential has been frequently used as an alternative to Renyi entropy [13].

A natural and important question now arises: what is the optimal estimate under the generalized EE risk (6)? We do not know the answer to this question in the general case. In this work, however, we will extend the results by Chen and Geman [2] to a more general case and show that, if the conditional density $p(x|y)$ is CSUM, the generalized MEE estimate will also be the conditional median (or equivalently, the conditional mean or mode).

---

[††] This quantity is called *information potential* since each term in its kernel estimator can be interpreted as a potential between two particles (see [13] for the physical interpretation of kernel estimator of information potential).

## II. MAIN THEOREM AND THE PROOF

In this section, the discussion is focused on the $\alpha$-order information potential (IP), but the conclusions drawn can be immediately transferred to Renyi entropy. The main theorem of the paper is as follows.

*Theorem1:* Assume for every value $y \in \mathbb{R}^m$, that the conditional PDF $p(x|y)$ is conditionally symmetric (rotation invariant for multivariate case) and unimodal (CSUM) in $x \in \mathbb{R}^n$, and let $\mu(y) = \text{median}\left[p(.|y)\right]$. If $\alpha$-order information potential $V_\alpha(X - \mu(Y)) < \infty$ ($\alpha > 0$, $\alpha \neq 1$), then

$$\begin{cases} V_\alpha(X - \mu(Y)) \leq V_\alpha(X - g(Y)) & \text{if } 0 < \alpha < 1 \\ V_\alpha(X - \mu(Y)) \geq V_\alpha(X - g(Y)) & \text{if } \alpha > 1 \end{cases} \quad (7)$$

for all $g: \mathbb{R}^m \to \mathbb{R}^n$ for which $V_\alpha(X - g(Y)) < \infty$.

*Remark*: As $p(x|y)$ is CSUM, the conditional median $\mu(y)$ in Theorem 1 is the same as the conditional mean $\pi(y)$ and conditional mode $\zeta(y)$. According to the relationship between information potential and Renyi entropy, the inequalities in (7) are equivalent to

$$H_\alpha\left(X - \mu(Y)\right) \leq H_\alpha\left(X - g(Y)\right) \quad (8)$$

*Proof of the Theorem*: In this work, we give a proof for the univariate case ($n = 1$). A similar proof can be easily extended to the multivariate case ($n > 1$). In the proof we assume, without loss of generality, that $\forall y$, $p(x|y)$ has median at $x = 0$, since otherwise we could replace $p(x|y)$ by $p(x + \mu(y)|y)$ and work instead with conditional densities centered at $x = 0$. The road map of the proof is similar to that contained in [2]. First, we prove the following proposition:

*Proposition1:* Assume that $f(x|y)$ (not necessarily a conditional density function) satisfies

1) non-negative, continuous and integrable in $x$ for each $y \in \mathbb{R}^m$;

2) symmetric (rotation invariant for $n > 1$) around $x = 0$ and unimodal for each $y \in \mathbb{R}^m$;

3) uniformly bounded in $(x, y)$;

4) $V_\alpha(f^0) < \infty$, where $V_\alpha(f^0) = \int_\mathbb{R} (f^0(x))^\alpha dx$, and $f^0(x) = \int_{\mathbb{R}^m} f(x|y)dF(y)$.

Then for all $g : \mathbb{R}^m \to \mathbb{R}$ for which $V_\alpha(f^g) < \infty$, we have

$$\begin{cases} V_\alpha(f^0) \leq V_\alpha(f^g) & \text{if } 0 < \alpha < 1 \\ V_\alpha(f^0) \geq V_\alpha(f^g) & \text{if } \alpha > 1 \end{cases} \quad (9)$$

where $V_\alpha(f^g) = \int_\mathbb{R} (f^g(x))^\alpha dx$, $f^g(x) = \int_{\mathbb{R}^m} f(x+g(y)|y)dF(y)$.

*Remark*: It is easy to observe that $\int f^0 dx = \int f^g dx \leq \sup_{(x,y)} f(x|y) < \infty$ (not necessarily $\int f^0 dx = 1$).

*Proof of the Proposition*: The proof is based on the following three lemmas.

*Lemma 1[2]*: Let non-negative function $h: \mathbb{R} \to [0,\infty)$ be bounded, continuous, and integrable, and define function $O_h(z)$ by

$$O_h(z) = \lambda\{x : h(x) \geq z\} \quad (10)$$

where $\lambda$ is Lebesgue measure. Then the following results hold:

a) Define $m^h(x) = \sup\{z : O_h(z) \geq x\}$, $x \in (0,\infty)$, and $m^h(0) = \sup_x h(x)$. Then $m^h(x)$ is continuous and non-increasing on $[0,\infty)$, and $m^h(x) \to 0$ as $x \to \infty$.

b) For any function $G : [0,\infty) \to \mathbb{R}$ with $\int_\mathbb{R} |G(h(x))| dx < \infty$

$$\int_\mathbb{R} G(h(x))dx = \int_0^\infty G(m^h(x))dx \quad (11)$$

c) For any $x_0 \in [0,\infty)$

$$\int_0^{x_0} m^h(x)dx = \sup_{A:\lambda(A)=x_0} \int_A h(x)dx \tag{12}$$

*Proof of Lemma 1*: See the proof of Lemma 1 in [2].

*Remark*: The transformation $h \to m^h$ in Lemma1 is also called the "rearrangement" of $h$ [7]. By Lemma 1, we have $V_\alpha(m^{f^g}) = V_\alpha(f^g) < \infty$ and $V_\alpha(m^{f^0}) = V_\alpha(f^0) < \infty$ (let $G(x) = x^\alpha$). Therefore, to prove Proposition 1, it suffices to prove

$$\begin{cases} V_\alpha(m^{f^0}) \leq V_\alpha(m^{f^g}) & \text{if } 0 < \alpha < 1 \\ V_\alpha(m^{f^0}) \geq V_\alpha(m^{f^g}) & \text{if } \alpha > 1 \end{cases} \tag{13}$$

*Lemma 2*: Denote $m^g = m^{f^g}$, $m^0 = m^{f^0}$. Then

a)

$$\int_0^\infty m^g(x)dx = \int_0^\infty m^0(x)dx < \infty \tag{14}$$

b)

$$\int_0^{x_0} m^g(x)dx \leq \int_0^{x_0} m^0(x)dx, \quad \forall x_0 \in [0,\infty) \tag{15}$$

*Proof of Lemma 2*: See the proof of Lemma 3 in [2].

*Lemma 3*: $\forall \alpha > 0$, let $n$ be a non-negative integer such that $n < \alpha \leq n+1$. Then $\forall x_0 \in [0,\infty)$

a)

$$\int_0^{x_0} \left\{ \left(m^g(x)\right)^{\alpha-n} \left(m^0(x)\right)^{n+1-\alpha} \right\} dx \leq \int_0^{x_0} m^0(x)dx \tag{16}$$

b)

$$\int_{x_0}^\infty \left\{ \left(m^g(x)\right)^{\alpha-n} \left(m^0(x)\right)^{n+1-\alpha} \right\} dx \leq \int_{x_0}^\infty m^g(x)dx \tag{17}$$

*Proof of Lemma 3*: According to Holder inequality [7], we have $\forall \Omega \subset [0,\infty)$,

$$\int_\Omega \left\{ \left(m^g(x)\right)^{\alpha-n} \left(m^0(x)\right)^{n+1-\alpha} \right\} dx \leq \left(\int_\Omega m^g(x)dx\right)^{\alpha-n} \left(\int_\Omega m^0(x)dx\right)^{n+1-\alpha} \tag{18}$$

By Lemma 2, $\int_0^{x_0} m^g(x)dx \leq \int_0^{x_0} m^0(x)dx$, it follows that

$$\int_0^{x_0} \left\{ \left(m^g(x)\right)^{\alpha-n} \left(m^0(x)\right)^{n+1-\alpha} \right\} dx \leq \left(\int_0^{x_0} m^g(x)dx\right)^{\alpha-n} \left(\int_0^{x_0} m^0(x)dx\right)^{n+1-\alpha}$$
$$\leq \left(\int_0^{x_0} m^0(x)dx\right)^{\alpha-n} \left(\int_0^{x_0} m^0(x)dx\right)^{n+1-\alpha} \tag{19}$$
$$= \int_0^{x_0} m^0(x)dx$$

Further, since $\int_0^\infty m^g(x)dx = \int_0^\infty m^0(x)dx$, we have $\int_{x_0}^\infty m^g(x)dx \geq \int_{x_0}^\infty m^0(x)dx$, and hence

$$\int_{x_0}^\infty \left\{ \left(m^g(x)\right)^{\alpha-n} \left(m^0(x)\right)^{n+1-\alpha} \right\} dx \leq \left(\int_{x_0}^\infty m^g(x)dx\right)^{\alpha-n} \left(\int_{x_0}^\infty m^0(x)dx\right)^{n+1-\alpha}$$
$$\leq \left(\int_{x_0}^\infty m^g(x)dx\right)^{\alpha-n} \left(\int_{x_0}^\infty m^g(x)dx\right)^{n+1-\alpha} \tag{20}$$
$$= \int_{x_0}^\infty m^g(x)dx$$

Q.E.D. (Lemma 3)

Let $S_g = \sup\{x : m^g(x) > 0\}$, which is finite or infinite, (17) can be rewritten as

$$\int_{x_0}^{S_g} \left\{ \left(m^g(x)\right)^{\alpha-n} \left(m^0(x)\right)^{n+1-\alpha} \right\} dx \leq \int_{x_0}^{S_g} m^g(x)dx \tag{21}$$

Now we are in position to prove (13):

(1) $0 < \alpha < 1$: In this case, we have

$$\int_0^\infty \left(m^g(x)\right)^\alpha dx = \int_0^{S_g} m^g(x)\left(m^g(x)\right)^{\alpha-1} dx$$

$$= \int_0^{S_g} m^g(x)\left(\int_0^{\left(m^g(x)\right)^{\alpha-1}} dy\right) dx$$

$$= \int_0^{S_g} \left\{\int_0^\infty m^g(x)\mathbb{I}\left(y \leq \left(m^g(x)\right)^{\alpha-1}\right) dy\right\} dx$$

$$= \int_0^\infty \left(\int_{\inf\left\{x:\left(m^g(x)\right)^{\alpha-1}\geq y\right\}}^{S_g} m^g(x) dx\right) dy$$

$$\overset{(A)}{\geq} \int_0^\infty \left\{\int_{\inf\left\{x:\left(m^g(x)\right)^{\alpha-1}\geq y\right\}}^{S_g} \left(m^g(x)\right)^{1-\alpha}\left(m^0(x)\right)^\alpha dx\right\} dy$$

$$= \int_0^\infty \left\{\int_0^{S_g} \left(m^g(x)\right)^{1-\alpha}\left(m^0(x)\right)^\alpha \mathbb{I}\left(\left(m^g(x)\right)^{\alpha-1}\geq y\right) dx\right\} dy$$

$$= \int_0^{S_g} \left(m^g(x)\right)^{1-\alpha}\left(m^0(x)\right)^\alpha \left(\int_0^{\left(m^g(x)\right)^{\alpha-1}} dy\right) dx$$

$$= \int_0^{S_g} \left(m^0(x)\right)^\alpha dx$$

$$\overset{(B)}{=} \int_0^{S_g} \left(m^0(x)\right)^\alpha dx + \int_{S_g}^\infty \left(m^0(x)\right)^\alpha dx \qquad (22)$$

$$= \int_0^\infty \left(m^0(x)\right)^\alpha dx$$

where (A) follows from (21), and (B) is due to $\int_{S_g}^\infty \left(m^0(x)\right)^\alpha dx = 0$, since

$$0 = \int_{S_g}^\infty m^g(x)dx \geq \int_{S_g}^\infty m^0(x)dx \geq 0$$
$$\Rightarrow \int_{S_g}^\infty m^0(x)dx = 0 \qquad (23)$$

(2) $\alpha > 1$: First we have

$$\int_0^\infty \left(m^0(x)\right)^\alpha dx = \int_0^\infty m^0(x)\left(m^0(x)\right)^{\alpha-1} dx$$

$$= \int_0^\infty m^0(x)\left(\int_0^{\left(m^0(x)\right)^{\alpha-1}} dy\right) dx$$

$$= \int_0^\infty \left\{\int_0^\infty m^0(x) \mathbb{I}\left(y \leq \left(m^0(x)\right)^{\alpha-1}\right) dy\right\} dx$$

$$= \int_0^\infty \left(\int_0^{\sup\left\{x:\left(m^0(x)\right)^{\alpha-1} \geq y\right\}} m^0(x) dx\right) dy$$

$$\stackrel{(C)}{\geq} \int_0^\infty \left\{\int_0^{\sup\left\{x:\left(m^0(x)\right)^{\alpha-1} \geq y\right\}} \left(m^g(x)\right)^{\alpha-n} \left(m^0(x)\right)^{n+1-\alpha} dx\right\} dy$$

$$= \int_0^\infty \left\{\int_0^\infty \left(m^g(x)\right)^{\alpha-n} \left(m^0(x)\right)^{n+1-\alpha} \mathbb{I}\left(\left(m^0(x)\right)^{\alpha-1} \geq y\right) dx\right\} dy$$

$$= \int_0^\infty \left(m^g(x)\right)^{\alpha-n} \left(m^0(x)\right)^{n+1-\alpha} \left(\int_0^{\left(m^0(x)\right)^{\alpha-1}} dy\right) dx$$

$$= \int_0^\infty \left(m^0(x)\right)^n \left(m^g(x)\right)^{\alpha-n} dx \qquad (24)$$

where (C) follows from (16). Further, one can derive

$$\int_0^\infty \left(m^0(x)\right)^n \left(m^g(x)\right)^{\alpha-n} dx = \int_0^\infty \left(m^0(x) \int_0^{\left(m^g(x)\right)^{\alpha-n} \left(m^0(x)\right)^{n-1}} dy\right) dx$$

$$= \int_0^\infty \left\{\int_0^\infty m^0(x) \mathbb{I}\left(y \leq \left(m^g(x)\right)^{\alpha-n} \left(m^0(x)\right)^{n-1}\right) dy\right\} dx$$

$$= \int_0^\infty \left\{\int_0^{\sup\left\{x:\left(m^g(x)\right)^{\alpha-n}\left(m^0(x)\right)^{n-1} \geq y\right\}} m^0(x) dx\right\} dy$$

$$\stackrel{(D)}{\geq} \int_0^\infty \left\{\int_0^{\sup\left\{x:\left(m^g(x)\right)^{\alpha-n}\left(m^0(x)\right)^{n-1} \geq y\right\}} m^g(x) dx\right\} dy$$

$$= \int_0^\infty \left\{\int_0^\infty m^g(x) \mathbb{I}\left(\left(m^g(x)\right)^{\alpha-n} \left(m^0(x)\right)^{n-1} \geq y\right) dx\right\} dy$$

$$= \int_0^\infty \left(m^0(x)\right)^{n-1} \left(m^g(x)\right)^{\alpha-n+1} dx$$

$$\vdots$$

$$\geq \int_0^\infty \left(m^0(x)\right)^{n-2} \left(m^g(x)\right)^{\alpha-n+2} dx$$

$$\vdots$$

$$\geq \int_0^\infty \left(m^g(x)\right)^\alpha dx \qquad (25)$$

12where (D) is because $\int_0^{x_0} m^g(x)dx \leq \int_0^{x_0} m^0(x)dx$, $\forall x_0 \in [0,\infty)$. Combining (24) and (25), we get $\int_0^\infty \left(m^0(x)\right)^\alpha dx \geq \int_0^\infty \left(m^g(x)\right)^\alpha dx$ (i.e. $V_\alpha(m^0) \geq V_\alpha(m^g)$).

Up to now, the proof of Proposition 1 has been completed. Let us come back to the proof of Theorem 1. The remaining task is just to remove the conditions of continuity and uniform boundedness imposed in Proposition 1. This can be easily accomplished by approximating $p(x|y)$ by a sequence of functions $\{f_n(x|y)\}$, $n=1,2,\cdots$, which satisfy the conditions of Proposition 1. Then similar to [2], we define

$$f_n(x|y) = \begin{cases} n\int_x^{x+(1/n)} \min(n, p(z|y))dz & \forall x \in [0,\infty) \\ f_n(-x|y) & \forall x \in (-\infty, 0) \end{cases} \quad (26)$$

It is easy to verify that for each $n$, $f_n(x|y)$ satisfies all the conditions of Proposition 1. Here we only give the proof for condition 4). Let $f_n^0(x) = \int_{\mathbb{R}^m} f_n(x|y)dF(y)$, we have

$$\begin{aligned}
V_\alpha(f_n^0) &= \int_{\mathbb{R}} \left(f_n^0(x)\right)^\alpha dx \\
&= \int_{\mathbb{R}} \left(\int_{\mathbb{R}^m} f_n(x|y)dF(y)\right)^\alpha dx \\
&= 2\int_{\mathbb{R}_+} \left(\int_{\mathbb{R}^m} \left(n\int_x^{x+(1/n)} \min(n, p(z|y))dz\right)dF(y)\right)^\alpha dx \\
&\leq 2\int_{\mathbb{R}_+} \left(\int_{\mathbb{R}^m} \left(n\int_x^{x+(1/n)} \sup_{z\in[x,x+(1/n)]} p(z|y)dz\right)dF(y)\right)^\alpha dx \quad (27)\\
&\stackrel{(E)}{=} 2\int_{\mathbb{R}_+} \left(\int_{\mathbb{R}^m} \left(n\int_x^{x+(1/n)} p(x|y)dz\right)dF(y)\right)^\alpha dx \\
&= 2\int_{\mathbb{R}_+} \left(\int_{\mathbb{R}^m} p(x|y)dF(y)\right)^\alpha dx \\
&= V_\alpha(p^0) < \infty
\end{aligned}$$

where (E) comes from the fact that $\forall y$, $p(x|y)$ is non-increasing in $x$ over $[0,\infty)$, since it is CSUM.

According to Proposition 1, we have, for every $n$,

$$\begin{cases} V_\alpha(f_n^0) \leq V_\alpha(f_n^g) & \text{if } 0 < \alpha < 1 \\ V_\alpha(f_n^0) \geq V_\alpha(f_n^g) & \text{if } \alpha > 1 \end{cases} \tag{28}$$

where $f_n^g(x) = \int_{\mathbb{R}^m} f_n(x + g(y) | y) dF(y)$. In order to complete the proof of Theorem 1, we only need to show that $V_\alpha(f_n^0) \to V_\alpha(p^0)$, and $V_\alpha(f_n^g) \to V_\alpha(p^g)$. This can be proved by the dominated convergence theorem. Here we only show $V_\alpha(f_n^0) \to V_\alpha(p^0)$, the proof for $V_\alpha(f_n^g) \to V_\alpha(p^g)$ is identical. First, it is clear that $f_n^0(x) \leq p^0(x)$, $\forall x$, and hence $\left(f_n^0(x)\right)^\alpha \leq \left(p^0(x)\right)^\alpha$, $\forall x$. Also, we can derive

$$\int_{\mathbb{R}} \left| f_n^0(x) - p^0(x) \right| dx \to 0 \tag{29}$$

As $V_\alpha(p^0) = \int_{\mathbb{R}} \left(p^0(x)\right)^\alpha < \infty$, then by dominated convergence theorem, $V_\alpha(f_n^0) \to V_\alpha(p^0)$.

Q.E.D. (Theorem 1)

## III. CONCLUSION

The problem of determining a minimum-error-entropy (MEE) estimator is actually the problem of shifting the components of a mixture of the conditional PDF so as to minimize the entropy of the mixture. It has been proved in a recent paper that, if the conditional distribution is conditionally symmetric and unimodal (CSUM), the Shannon entropy of the mixture distribution will be minimized by aligning the conditional median. In this work, this result has been extended to a more general case. We show that if the conditional distribution is CSUM, the Renyi entropy of the mixture distribution will also be minimized by aligning the conditional median.

ACKNOWLEDGEMENTS

This work was supported by National Natural Science Foundation of China (No. 61372152), and 973 Programs (No. 2012CB316400, 2012CB316402).

REFERENCES

[1] R. K. Boel, M. R. James, I. R. Petersen, Robustness and risk-sensitive filtering, *IEEE Trans. Automatic Control*, vol. 47, no. 3, pp. 451-461, 2002.

[2] T. L. Chen, S. Geman, On the minimum entropy of a mixture of unimodal and symmetric distributions. *IEEE Transactions on Information Theory*, vol. 54, no.7, July 2008, pp. 3166-3174.

[3] T. M. Cover and J. A. Thomas, *Element of Information Theory,* Chichester: Wiley & Son, Inc., 1991.

[4] D. Erdogmus, J.C. Principe, Generalized information potential criterion for adaptive system training, *IEEE Trans. on Neural Networks,* vol. 13, 2002, pp. 1035-1044.

[5] D. Erdogmus, J.C. Principe, From linear adaptive filtering to nonlinear information processing, *IEEE Signal Processing Magazine*, vol. 23, no. 6, 2006, pp. 15-33.

[6] E. B. Hall, G. L. Wise, On optimal estimation with respect to a large family of cost functions, *IEEE Trans. Inform. Theory*, vol. IT-37, no. 3, 1991, pp. 691-693.

[7] G. H. Hardy, J. E. Littlewood, G. Polya, *Inequalities*, Cambridge, U. K. : Cambridge Univ. Press, 1934.

[8] S. Haykin, *Adaptive Filtering Theory*, 3$^{rd}$ ed., NY: Prentice Hall, 1996.

[9] T. Kailath, B. Hassibi, *Linear Estimation*, Prentice Hall, NJ, 2000.

[10] P. Kalata, R. Priemer, Linear prediction, filtering and smoothing: an information theoretic approach, *Information Science*, vol. 17, pp. 1-14, 1979.

[11] A. Papoulis, S. U. Pillai, *Probability, Random Variables, and Stochastic Processes*, 4$^{th}$ Edition. McGraw-Hill Companies, Inc., 2002.

[12] S. C. Pei, C. C. Tseng, Least mean p-power error criterion for adaptive FIR filter, *IEEE Journal on Selected Areas in Communications*, vol. 12, no. 9, 1994, pp. 1540-1547.

[13] J. C. Principe, *Information Theoretic Learning: Renyi's Entropy and Kernel Perspectives*. New York: Springer, 2010.

[14] A. Renyi, On measures of entropy and information, *Proc. of the Fourth Berkeley Symposium on Mathematical Statistics and Probability*, vol. 1, 547-561, 1961.

[15] P. J. Rousseeuw, A. M. Leroy, *Robust Regression and Outlier Detection*, New York: John Wiley & Sons, Inc., 1987.

[16] M. Shao, C. L. Nikias, Signal processing with fractional lower order moments: stable processes and their applications, *Proceedings of the IEEE*, vol. 81, no. 7, 1993, pp. 986-1009.

[17] Y. Tomita, S. Ohmatsu, T. Soeda, An application of the information theory to estimation problems, *Information and Control*, 1976, 32:101–111.

[18] H. L. Weidemann, E. B. Stear, Entropy analysis of estimating systems, *IEEE Transactions on Information Theory*, vol. 16, no. 3, pp.264-270, 1970.